\documentclass[runningheads]{llncs}
\usepackage{llncsdoc}

\usepackage{amsmath}
\usepackage{amssymb}
\usepackage{graphicx}
\usepackage{amsfonts}
\usepackage{bm}
\usepackage{multirow}
\usepackage{caption}
\usepackage{subfig}
\usepackage{hyperref}

\usepackage{color}
\definecolor{light-gray}{RGB}{224,224,224}
\definecolor{gray}{RGB}{190,190,190}

\DeclareMathOperator*{\argmax}{arg\,max}

\usepackage{url}
\urldef{\mailsa}\path|ymkim@kisti.re.kr|
\urldef{\mailsb}\path|{julien.velcin,stephane.bonnevay,marian-andrei.rizoiu}@univ-lyon2.fr|
\newcommand{\keywords}[1]{\par\addvspace\baselineskip
\noindent\keywordname\enspace\ignorespaces#1}

\begin{document}

\mainmatter  

\title{Temporal Multinomial Mixture for Instance-oriented Evolutionary Clustering}

\titlerunning{Temporal Multinomial Mixture}

%
%
\author{Young-Min Kim$^{\dagger}$%
\and Julien Velcin$^{\ddagger}$\and St\'ephane Bonnevay$^{\ddagger}$ \and\\ Marian-Andrei Rizoiu$^{\ddagger}$}
\authorrunning{Temporal Multinomial Mixture}

\institute{$^{\dagger}$Korea Institute of Science and Technology Information, South Korea\\
$^{\ddagger}$ERIC Lab., University of Lyon 2, France\\
\mailsa\\
\mailsb\\
}

%
%

\toctitle{Lecture Notes in Computer Science}
\tocauthor{Authors' Instructions}
\maketitle

\begin{abstract}
Evolutionary clustering aims at capturing the temporal evolution of clusters.
This issue is particularly important in the context of social media data that are naturally temporally driven.
In this paper, we propose a new probabilistic model-based evolutionary clustering technique.
The Temporal Multinomial Mixture (TMM) is an extension of classical mixture model that optimizes feature co-occurrences in the trade-off with temporal smoothness.
Our model is evaluated for two recent case studies on opinion aggregation over time.
We compare four different probabilistic clustering models and we show the superiority of our proposal in the task of instance-oriented clustering.
\end{abstract}

\section{Introduction} 
\label{sec:intro}

Clustering is a popular way to preprocess large amount of unstructured data.
It can be used in several ways, such as data summarization for decision making or representation learning for classification purpose.
Recently, evolutionary clustering aims at capturing temporal evolution of clusters in data streams.
This is different from traditional incremental clustering, for evolutionary clustering methods optimize another measure that builds the clustering model at time $t+1$ by taking into account of the model at time $t$ in a retrospective manner \cite{Chakrabarti:2006,Chi:2007,Xu:2008b}.
Applications range from clustering photo tags in \texttt{flickr.com} to document clustering in textual corpora.

The existing methods fall into two different categories.
\emph{Instance-oriented} evolutionary clustering mostly aims at primarily regrouping objects and \emph{topic-oriented} evolutionary clustering aims at estimating distributions over components (\textit{e.g.}, words).
While the former extracts tightest clusters in the feature space, the latter improves the smoothness of temporally consecutive clusters.
In this work, we focus on developing a new temporal-driven model of the first category, motivated by two case studies.

We propose a new probabilistic evolutionary clustering method that aims at finding dynamic instance clusters.
Our model, Temporal Mixture Model (TMM), is an extension of the classical mixture model to categorical data streams.
The main novelty is not to use Dirichlet prior in order to relax smoothness constraint.
While our model can further be improved in terms of more advanced properties, such as learning the number of clusters as in non-parametric models \cite{Teh:2006,Ahmed:2008}, in this work we mainly focus on realizing our basic idea and studying the performance of the model. Using internal evaluation measures, we demonstrate that TMM outperforms a typical \emph{topic-oriented} dynamic model and achieves similar compactness results with two static models.
This result is achieved at the slight expense of cluster smoothing ability through temporal epochs.

In the following sections, we first motivate and present in detail the proposed TMM model.
Then we present the experimental results of TMM as well as three other methods of the literature, showing the superiority of our method with new type of datasets in opinion mining.
Finally we conclude with some perspectives and future works.

\section{Motivation and related work} 

\subsection{Motivation}

Document clustering and topic extraction are sometimes considered as equivalent problems, and the methods desired to address each problem are used interchangeably \cite{Zhang:2010}. However, there is a fundamental difference in terms of clustering objective between them and this draws a clear algorithmic difference. 
Even though this issue has not been actively mentioned in the clustering literature, it is indirectly confirmed by the fact that topic modeling is not recommended to be used directly for document clustering in general. 
\cite{Pessiot:2010} have empirically shown that even simple mixture models outperform Dirichlet distribution-based topic models for document clustering, when directly using model parameters. 
A recent work \cite{Xie:2013} is dealing with this issue by proposing an integrated graphical model for both document clustering and topic modeling. 
However, the great success of topic models in unsupervised learning has often led researchers to use them as instance clustering in practice.
This observation remains valid for evolutionary clustering, for which one hardly finds an alternative to topic models using Dirichlet smoothing.
The situation is identical when dealing with more classical categorical data, which is the case of our work.
This paper starts from this significant issue in evolutionary clustering. 

To the best of our knowledge, this is the first attempt to use a non-Dirichlet mixture model for temporal analysis of data steams. 
The reason why we abandon Dirichlet prior reflects our (maybe peculiar) point of view towards the Dirichlet distribution. 
That is, the power of topic models mainly comes from their ability to smoothen distributions via the Dirichlet prior. 
It is effective for extracting representative topics or for making inference on new data. 
However, in case of clustering instances, a hasty smoothing of the distributions risks to mix data samples with no common feature. 
In this paper, target datasets are not necessarily textual; therefore the clustering process can be more sensitive to this effect than when dealing with a large feature space (such as a vocabulary of words).
In our case, each feature becomes more important, thus special attention must be given to the actual matching between the cluster distribution and the observed feature co-occurrences.
This is the reason why we decide to build our method on top of a simple mixture model expecting to minimize the discussed risk. 

\subsection{Related work}

\begin{figure}[t]
	\centering
  	\subfloat[] {
		\includegraphics[scale=0.32]{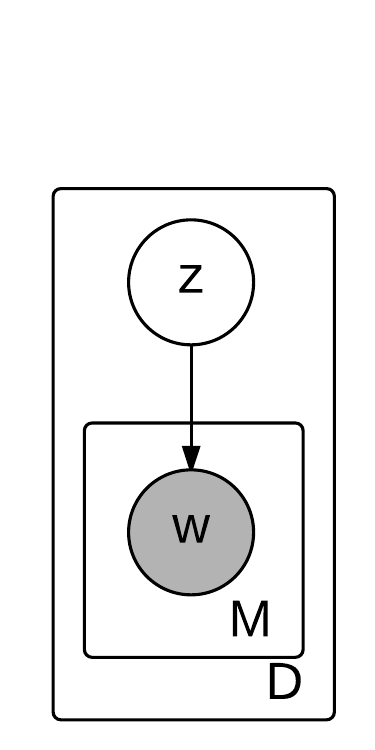}
	}
    \hspace{0.03\textwidth}
    \subfloat[] {
		\includegraphics[scale=0.32]{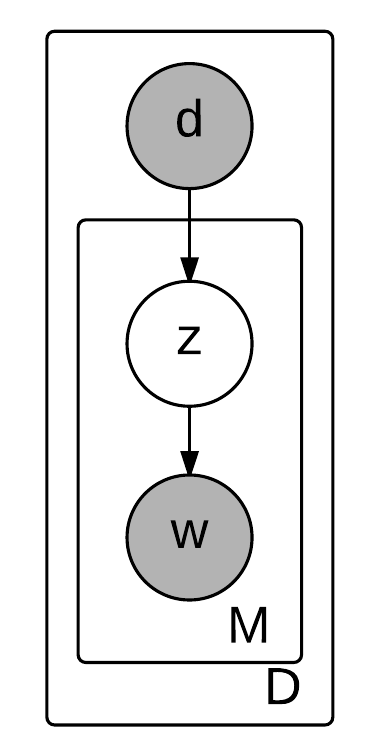}
	}
    \hspace{0.03\textwidth}
    \subfloat[] {
		\includegraphics[scale=0.32]{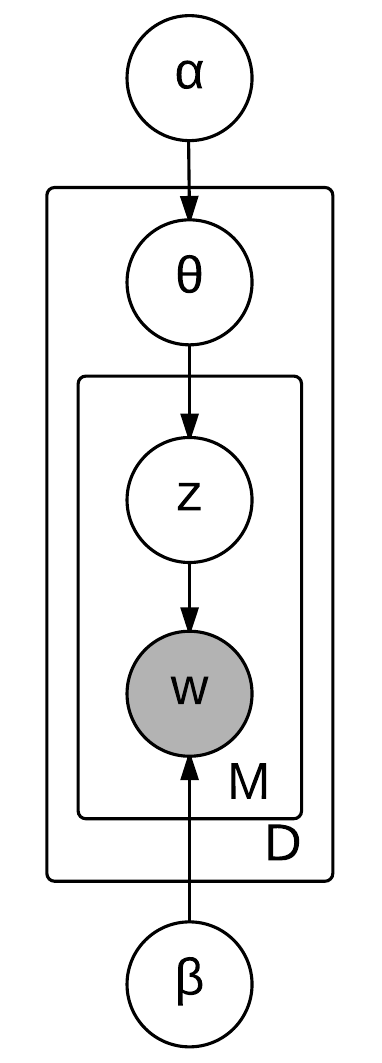}
	}
    \hspace{0.03\textwidth}
    \subfloat[] {
		\includegraphics[scale=0.32]{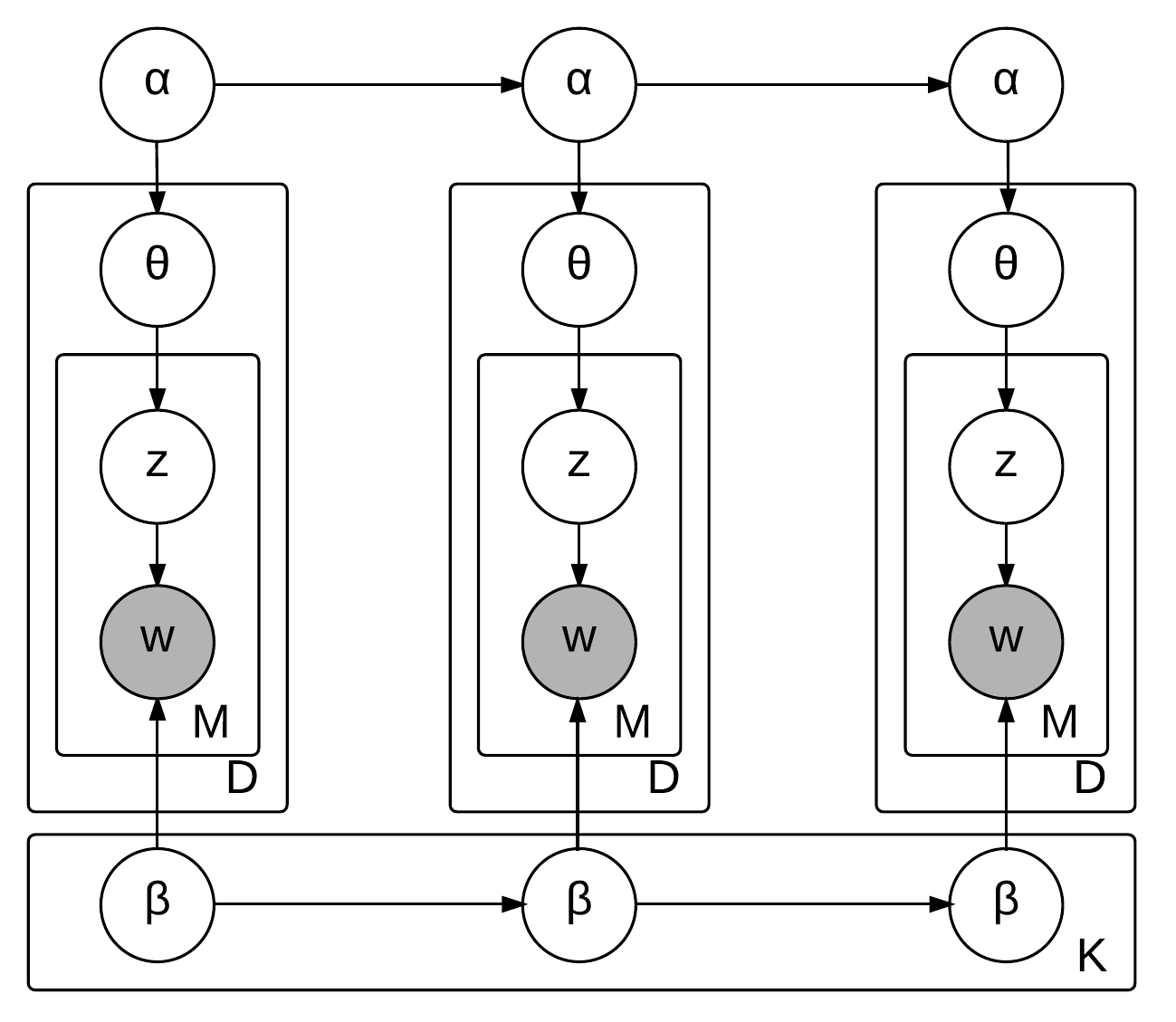}
	}
    \caption{Graphical representation of (a) MM, (b) PLSA, (c) LDA, and (d) DTM.}
    \vspace{-0.5cm}
  \label{fig:generative}
\end{figure}

Our new evolutionary clustering model, \emph{Temporal Multinomial Mixture} (TMM), has been designed with the assumption that regrouping non co-occurring features is highly prejudicial.
TMM is a temporal extension of the \emph{Multinomial Mixture} (MM), a simple probabilistic generative model for clustering. 
More complex mixture models such as \emph{Probabilistic Latent Semantic Analysis} (PLSA) \cite{Hofmann:1999} or \emph{Latent Dirichlet Allocation} (LDA) \cite{Blei:2003} seem less suitable for clustering non-textual data as mentioned in Section 2.1.
Non co-occurring features are often mixed together in the same cluster because of additional hidden layers added to these models, either for instance-topic distributions (PLSA) or as Dirichlet prior (LDA).
The graphical representation of these models are given in Fig.~\ref{fig:generative}(a)-(c).  

Despite the obvious difference between our purpose and dynamic topic models, since the temporal approaches in unsupervised learning usually stand on the basis of topic models, it is inevitable to introduce the state-of-the arts of topic models. 
Most of the current techniques in clustering introducing a temporal dimension are topic models taking Dirichlet distribution \cite{Wei:2007,Iwata:2010} since the development of \emph{Dynamic Topic Model} (DTM, Fig.~\ref{fig:generative}(d)) \cite{Blei:2006}, a simple extension of LDA. This kind of dynamic topic analysis has been the object of numerous studies over recent years 
and more complex models such as DMM \cite{Wei:2007} or MDTM \cite{Iwata:2010} have been developed.
In comparison, TMM is much simpler and we experimentally show the power of simple modeling by comparing three clustering methods, MM, PLSA and DTM with ours.

On the other hand, some pioneer works were designed for data points that basically last during more than two time periods.
These stand on various theoretical bases such as k-means, agglomerative hierarchical method, spectral clustering, and even generative model \cite{Chakrabarti:2006,Chi:2007,Lin:2008}. However, the underlined property of data points is contrary to the case of data stream, which is our concern here. 

Whatsoever, several applications in temporal analysis are intended for dealing with text corpora.
Being designed for text hinders the ``out-of-the-box'' application of these methods to unfamiliar data such as image, gene, market, network data etc. 
In comparison, TMM is an evolutionary clustering dedicated to general categorical datasets.

\section{Temporal Multinomial Mixture}
\label{sec:tmm}
We propose Temporal Multinomial Mixture (TMM) for instance-oriented clustering over time. 
TMM is a temporal extension of MM and the relation between TMM and MM is analogous to that between DTM and LDA.
While the majority of existing temporal topic analysis tend to complicate the modeling process, TMM rather goes against this trend. 
We assume that complicated distributional structures confuse the instance-oriented clustering. 
Therefore our method assumes the form of a simple mixture model.
As in many other evolutionary clusterings and temporal topic analysis, data instances are associated with a time epoch. 
A time epoch indicates a time period between two adjacent moments. 
Dataset is generally divided into subsets by epoch. 
Instances are assumed to be described by features weighted with a frequency\footnote{For the sake of understanding, the reader can see a feature as a unique word over a vocabulary and a data component as a word occurrence in a document even if an instance is not a document here.}.

\subsection{Generative process}
\begin{figure}[htb]
  \centering
  \vspace{-0.5cm}
   \includegraphics[scale=0.43]{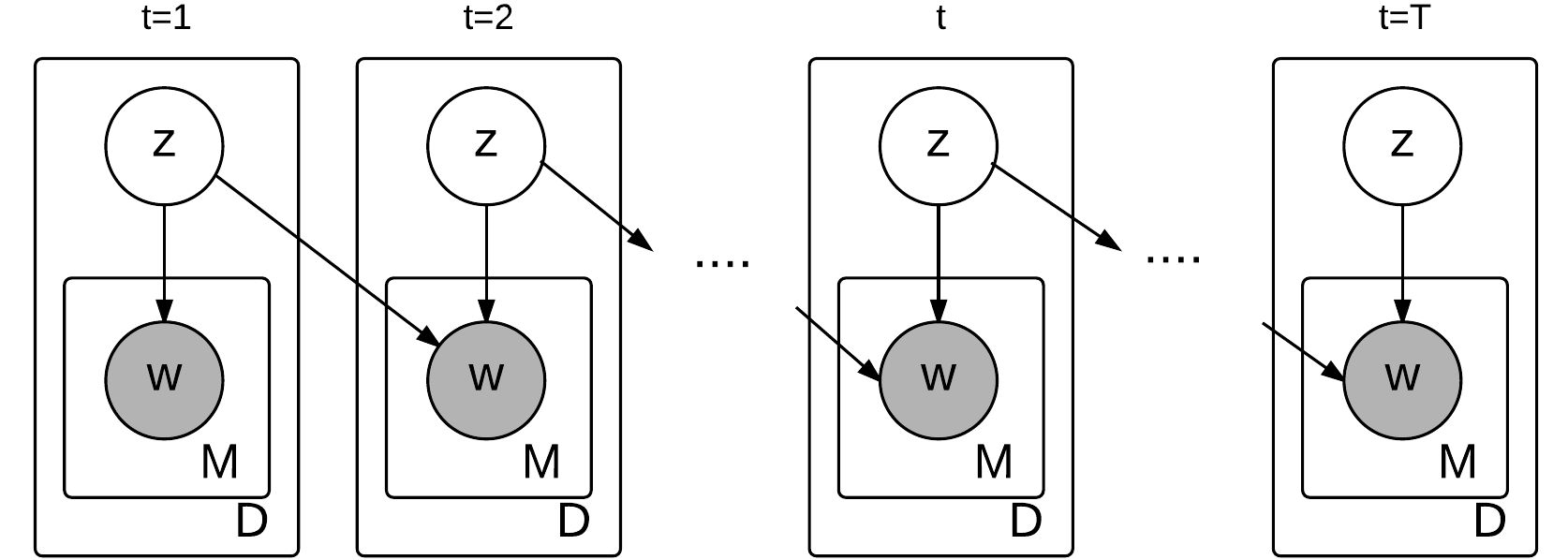}
  \caption{Graphical representation of a temporal multinomial mixture model}
  \vspace{-0.5cm}
  \label{fig:tmm}
\end{figure}

The graphical representation of TMM is given in Fig.~\ref{fig:tmm}. 
The extension from MM is realized by encoding the temporal dependency into the relation between data components $w$ of the current epoch and the clusters $z$ of the previous epoch.
The generation process of an instance $d^t=i$ at the epoch $t$ is as follows:
\begin{itemize}
\item Choose a cluster $z^{t-1}_i$ with probability $p(z^{t-1}_i)$.
\item Choose a cluster $z^{t}_i$ with probability $p(z^{t}_i)$.
\item Generate an instance $d^t=i$ with probability  $p(d^t=i | z^{t-1}_i, z^{t}_i)$  when $t>1$ or with $p(d^1=i | z^{1}_i)$ when $t=1$.
\end{itemize}
The last step is realized by repeatedly generating the components $w^t_{im}, \forall m$, sequential features in the instance $d^t=i$, as illustrated in the graphical representation. 
Unlike most temporal graphical models, it is a connected network considering the correlation of all topics of $t$ and $t-1$.
The notations used in TMM are shown in Table \ref{tab:symbol}. We mostly referred the notations in \cite{AlSumait:2008} and \cite{He:2013}.
Because of the variable dependency between different time epochs, we need sequential expression of features.
This is the reason why we cannot use the simple notation of MM.

\begin{table}[t!]
\centering
\scriptsize
\caption{Notations}
  \label{tab:symbol}
  \begin{tabular}{ lp{0.85\textwidth} }
  \hline
  Symbol & Description \\\hline
  $d^t$ & instance $d$ at epoch $t$\\
  $w^t_{im}$ & $m$th component in the instance  \text{$\setlength{\thickmuskip}{0mu}d^t=i$ at epoch $t$} \\
  $z^t_{i}$ & assigned cluster for instance \text{$\setlength{\thickmuskip}{0mu}d^t=i$ at epoch $t$} \\
  $D^t$ & sequence of instances at epoch $t$\\
  $Z^t$ & sequence of cluster assignments for $D^t$\\
  $\mathbf{D}$ & sequence of all instances, $\mathbf{D} = (D^1, D^2, ..., D^T)$\\
  $\mathbf{Z}$ & sequence of cluster assignments for $\mathbf{D}$,  $\mathbf{Z} = (Z^1, Z^2, ..., Z^T)$\\
  $T$ & number of epochs\\
  $|D^t|$ & number of instances at epoch $t$\\
  $M^t_{d}$ & number of components in the instance $d$ at epoch $t$\\
  $V$ & number of unique components (number of features)\\
  $K$ & number of clusters\\
  $\bm{\phi}^t_k$ & multinomial distribution of cluster $k$ over components at epoch $t$ \\
  $\pi^t_k$ & prior probability of cluster $k$ at epoch $t$ \\
  $\alpha$ & weight for the component generation from the clusters of previous epoch, $0 < \alpha < 1$ \\\hline
  \end{tabular}
  \vspace{-0.5cm}
\end{table}

\subsection{Parameter estimation via approximate development}
The objective function to be maximized is the expectation of log-likelihood \cite{Bishop:2006}:
\begin{equation}
	\label{eq:EL}
	\text{\small $ \mathbb{E}(\tilde{\mathcal{L}}) = \sum_{\mathbf{Z}} p(\mathbf{Z}|\mathbf{D}, \Theta_{old}) \cdot \log \left(p(\mathbf{D,Z}|\Theta)\right)$}
\end{equation}
Because of the dependency between the variables $z^t$ and $z^{t-1}$, the log-likelihood cannot be simplified using marginalized latent variables as in MM or PLSA. 
Instead, we start with the joint distribution of instances and assigned clusters (latent variables):
\begin{equation}
  \label{eq:joint}
	\text{\small $ p(\mathbf{D,Z}) = \left\{ \prod^{|D^1|}_{d=1} p(z^1_{d})\cdot p(d^1|z^1_{d}) \right\} \left\{ \prod^{T}_{t=2}\prod^{|D^t|}_{d=1} p(z^{t}_d) \cdot p(d^t|z^{t}_{d}, z^{t-1}_{d}) \right\}$}
\end{equation}
Eq. \ref{eq:EL} can be simplified by taking only the valid latent variables per term:
\begin{align}{}
	&\text{\small $ \mathbb{E}(\tilde{\mathcal{L}})	=   \sum_{i=1}^{|D^1|} \sum_{k=1}^{K} p(z^1_{i}=k | d^1=i) \log \{ p(z^1_{i}=k)p(d^1=i | z^1_{i}=k) \}$} \nonumber\\
	&\text{\small $ \setlength{\thickmuskip}{0mu} +  \sum_{t=2}^{T} \sum_{i=1}^{|D^t|} \sum_{k=1}^{K} \sum_{k'=1}^{K} p(z^t_{i}=k, z^{t-1}_{i}=k' | d^t=i) \log \{ p(z^t_{i}=k)p(d^t=i | z^t_{i}=k, z^{t-1}_{i}=k')  \}$}
\end{align}
At epoch $1$, $\setlength{\thickmuskip}{0mu}p(d^1=i | z^1_i=k)$ can be rewritten using $\phi^1_k$ and $n^1_{i,j}$, the frequency of unique component $j$ included in the instance $i$, such as {\small $\prod_{j=1}^{V}  {(\phi^1_{k,j})}^{n^1_{i,j}}$}.
On the other hand, the instance generation at epoch $t$, $\forall t\geq2$ is dependent also on the clusters of the previous epoch. 
Thus the conditional probability of an instance $i$ given current and previous clusters $k$ and $k'$, is inferred as follows with Bayes Rule:
\begin{align}{}
	\label{eq:bayes}
\text{$\small\setlength{\thickmuskip}{0mu}\setlength{\thinmuskip}{0mu}\setlength{\medmuskip}{0mu} p(d^t=i | z^t_i=k, z^{t-1}_i=k') = \prod_{m=1}^{M_{i}^{t}} \frac{p(z^t_{i}=k | w^t_{im}, z^{t-1}_i=k') p(z^{t-1}_i=k' | w^t_{im}) p(w^t_{im})}{p(z^t_{i}=k, z^{t-1}_i=k')}$}
\end{align}
Under the assumptions of graphical model, the analytical calculation of $\setlength{\thickmuskip}{0mu}p(z^t_{i} | w^t_{im},$ $z^{t-1}_i)$ is so complicated because the latent variables are related by the explaining away effect. To tackle this issue, we make an important hypothesis that $p(z^t_{i} | w^t_{im}, z^{t-1}_i)$ can be \textit{approximated} by $p(z^t_{i} | w^t_{im})$. 
Consequently, Eq. \ref{eq:bayes} is rewritten using  $\setlength{\thickmuskip}{0mu}p(w^t_{im}=j |z^{t}_i=k)$ as well as $\setlength{\thickmuskip}{0mu}p(w^t_{im}=j |z^{t-1}_i=k')$, which is equivalent to the previous epoch's parameter $\phi^{t-1}_{k',j}$. Penalizing the influence rate of the previous cluster with $\alpha$, a weighted parameter value $(\phi^{t-1}_{k',j})^{\alpha}, 0<\alpha<1$ is used instead of $\phi^{t-1}_{k',j}$. Letting the constant $\prod_{m=1}^{M_{i}^{t}}1/p(w^t_{im})$ be $C^t_i$, we obtain the following equation.
\begin{align}{}
	\label{eq:doc2word4}
	 p(d^t=i | z^t_i=k, z^{t-1}_i=k')
					& = C^t_i\cdot \prod_{j=1}^{V}  {(\phi^t_{k,j})}^{n^t_{i,j}} {(\phi^{t-1}_{k',j})}^{\alpha\cdot n^t_{i,j}}
\end{align}
Using the parameters $\Theta$, the $\mathbb{E}(\tilde{\mathcal{L}})$ becomes:
{\small
\begin{align}{}
	&\text{$\setlength{\thickmuskip}{0mu} \mathbb{E}(\tilde{\mathcal{L}})
							 =  \sum_{i=1}^{|D^1|} \sum_{k=1}^{K} p(z^1_{i}=k | d^1=i) \cdot \left\{ \log \pi^1_k + \sum_{j=1}^{V} {n^1_{i,j}} \cdot \log {\phi^1_{k,j}}  \right\}$} \nonumber\\
	&\text{$\setlength{\thickmuskip}{0mu}\setlength{\thinmuskip}{0mu}\setlength{\medmuskip}{0mu}   +  \sum_{t=2}^{T} \sum_{i=1}^{|D^t|} \sum_{k=1}^{K} \sum_{k'=1}^{K} p(z^t_{i}=k, z^{t-1}_{i}=k' | d^t=i) \cdot \left\{ \log \pi^t_{k} + \log C^t_i   + \sum_{j=1}^{V} n^t_{i,j} \cdot \left( \log {\phi^t_{k,j}} + \alpha\cdot \log {\phi^{t-1}_{k',j}} \right)  \right\} $}\nonumber
\end{align}
}

\subsection{EM algorithm}
We solve the following optimization problem to obtain the parameter values.
\begin{align*}{}
	\argmax_{\Theta} ~ \mathbb{E}(\tilde{\mathcal{L}}) , ~~
	\mathrm{subject~to}	 \sum_{j=1}^{V} \phi^t_{k,j} = 1, ~   \forall t,k ~~~ \mathrm{and} ~~~ \sum_{k=1}^{K}  \pi^t_k= 1, ~   \forall t.
\end{align*}
The EM algorithm is updated as follows.\vspace{0.3cm}\\
\underline{Initialization}\\
Randomly initialize parameters $\Theta = \{ \bm{\phi}^t_k, \pi^t_k ~|~ \forall t,k\}$
\begin{align*}{}
	\mathrm{subject~to}	~~ \sum_{j=1}^{V} \phi^t_{k,j} = 1, ~~   \forall t,k   ~~    ~~    \text{and}  ~~    ~~   \sum_{k=1}^{K}  \pi^t_k= 1, ~~  \forall t.
\end{align*}
\underline{E-step}\\
Compute the expectation of posteriors as follows.
\begin{align}
	&\text{\small$\setlength{\thickmuskip}{0mu} p(z^t_i=k, z^{t-1}_i=k' | d^t=i) = \frac{\prod_{j=1}^{V} {(\phi^t_{k,j})}^{n^t_{i,j}} {(\phi^{t-1}_{k',j})}^{\alpha\cdot n^t_{i,j}} \cdot  \pi^t_{k} \cdot \pi^{t-1}_{k'}}{\sum\limits_{a=1}^{K} \sum\limits_{a'=1}^{K} \prod_{j=1}^{V} {(\phi^t_{a,j})}^{n^t_{i,j}} {(\phi^{t-1}_{a',j})}^{\alpha\cdot n^t_{i,j}} \cdot  \pi^t_{a} \cdot \pi^{t-1}_{a'}}, 2 \leq t \leq T, \forall k,k',i.$}
\end{align}
{\small$p(z^1_i=k | d^1=i)$} is similarly calculated by eliminating the variables of $t-1$.\\\\
\underline{M-step}\\
Update the parameters maximizing the objective function.
\begin{align}
	\text{\small$\phi^t_{k,j}  =$}&\text{\small$\setlength{\thickmuskip}{0mu}\setlength{\thinmuskip}{0mu}\setlength{\medmuskip}{0mu} \frac{\sum\limits_{i=1}^{|D^t|} \sum\limits_{k'=1}^{K}  n^t_{i,j} \cdot p(z^t=k, z^{t-1}=k' | d^t=i)  + \sum\limits_{i=1}^{|D^{t+1}|} \sum\limits_{k'=1}^{K} \alpha\cdot n^{t+1}_{i,j} \cdot p(z^{t+1}_i=k', z^t_i=k | d^{t+1}=i) }{\sum\limits_{i=1}^{|D^t|}\sum\limits_{j'=1}^{V}\sum\limits_{k'=1}^{K} n^t_{i,j'} \cdot p(z^t=k, z^{t-1}=k' | d^t=i) + \sum\limits_{i=1}^{|D^{t+1}|} \sum\limits_{j'=1}^{V}\sum\limits_{k'=1}^{K} \alpha\cdot n^{t+1}_{i,j'} \cdot p(z^{t+1}_i=k', z^t_i=k | d^{t+1}=i)}$},\nonumber\\
	&\text{\small$~~2 \leq t \leq T-1, ~~ \forall j,k.$}
\end{align}
$\phi^1_{k,j}$ is calculated by eliminating the variables of $t-1$ from the above formula and $\phi^T_{k,j}$ is done by eliminating both variables and terms of $t+1$.
\begin{align}	
	&\text{\small$\setlength{\thickmuskip}{0mu}\setlength{\thinmuskip}{0mu}\setlength{\medmuskip}{0mu} \pi^t_k = \frac{\sum\limits_{i=1}^{|D^t|} \sum\limits_{k'=1}^{K}  p(z^t=k, z^{t-1}=k' | d^t=i)  + \sum\limits_{i=1}^{|D^{t+1}|} \sum\limits_{k'=1}^{K}  p(z^{t+1}_i=k', z^t_i=k | d^{t+1}=i) }{\sum\limits_{i=1}^{|D^t|}\sum\limits_{a=1}^{K}\sum\limits_{k'=1}^{K}  p(z^t=a, z^{t-1}=k' | d^t=i) + \sum\limits_{i=1}^{|D^{t+1}|} \sum\limits_{k'=1}^{K}\sum\limits_{a=1}^{K}  p(z^{t+1}_i=k', z^t_i=a | d^{t+1}=i)},$}\\
	&\text{\small$~~~~~~2 \leq t \leq T-1, ~~ \forall k$}\nonumber
\end{align}
$\pi^1_k $ and $\pi^T_k $ are calculated as in $\phi^1_{k,j}$ and $\phi^T_{k,j}$.

\subsection{Instance assignment and cluster evolution}
The assignment of each instance is eventually obtained from the estimated distributions.
For $t=1$, we assign to the instance $i$ the cluster that maximizes the posterior probability $\setlength{\thickmuskip}{0mu}p(z^1_i=k | d^1=i)$. 
For the instances in the other epochs, we integrate out $z^{t-1}_i$ to obtain the instance cluster such that $\setlength{\thickmuskip}{0mu}p(z^t_i=k | d^t=i) = \sum_{k'=1}^{K} p(z^t_i=k, z^{t-1}_i=k' | d^t=i).$

TMM being a connected network, all the clusters in the epoch $t-1$ can contribute to the clusters in the epoch $t$.
Please note that the same cluster index in different epochs does not mean that the corresponding clusters are identical over time.
That is why we need to find which cluster of the previous epoch contributes most to the specific cluster $k$ of the current epoch.
The dynamic correlation between clusters of the adjacent epochs is fully encoded in the distribution $\setlength{\thickmuskip}{0mu}p(z^t_i=k, z^{t-1}_i=k' | d^t=i)$.
By integrating out $z^{t}_i$ instead of $z^{t-1}_i$ from $\setlength{\thickmuskip}{0mu}p(z^t_i=k, z^{t-1}_i=k' | d^t=i)$, we can deduce the most likely cluster at the previous epoch for the instance $\setlength{\thickmuskip}{0mu}d^t=i$. We call it the origin of the instance. Given the specific cluster $z^t=k$, we have the classified instances and their origins.
By counting we find the most frequent origin and we can eventually relate the most influential cluster of the previous epoch to $z^t=k$.
Since this is a surjective function from $t$ to $t-1$, the division of a cluster over time is traceable.
Conversely, the merge of multiple clusters can also be caught if we choose not only the most likely cluster but also the second or the third likely one.

\section{Experiments}
\label{sec:exp}
We compare four different generative models in order to evaluate the performance of TMM. 
DTM is selected as a Dirichlet-based model; MM and PLSA are used as static baselines for highlighting the effect of introducing a temporal dimension.
Finally, we show that TMM outperforms the other models on two datasets of opinion mining, by finding a trade-off between compactness and temporal smoothing.

\subsection{Datasets}
\textit{\textbf{ImagiWeb political opinion dataset}}\footnote{It will be distributed to the public in Spring 2015 on the ImagiWeb official website, http://mediamining.univ-lyon2.fr/velcin/imagiweb.} 
The first dataset is comprised of a set of about 7000 unique tweets related to two politicians (each politician is analyzed separately).
The manual annotation process has been supervised by domain experts of public opinion analysis and it has followed a detailed procedure with the design of 9 aspects (\textit{e.g.}, project, ethic or political line) targeted by 6 possible opinion polarities (-2=very negative, -1=negative, 0=neutral, +1=positive, +2=very positive, NULL=ambiguous).
For instance, the tweet ``RT @anonym: P's project is just hot air'' can be described by the pair \texttt{(project,-2)} attached to the politician $P$.
Each pair corresponds to a feature $w$ whose value is the occurrence of the corresponding opinion for describing the studied entity. 
The full procedure and dataset are described in \cite{Velcin:2014}.
Because of the length limit of a tweet as well as for clustering purpose, we decide to combine the annotations by author for each time epoch.
 

 \medskip
\noindent
\textit{\textbf{RepLab 2013 Corpus}}
This corpus has been used for the RepLab 2013, second evaluation campaign on Online Reputation Management. It consists of a collection of tweets referring to 61 entities from four domains. We select two dominant domains out of four, automative and music, where the number of entities is 20 respectively. The clustering is done for each \textit{domain} separately this time instead of entity.
Tweets are annotated with three polarities: positive, negative and neutral. 
We let the features be the \textit{entity-polarity} pairs instead of aspect-polarity pairs, so that the opinion aggregation is based on co-occurring entities.
It means that the opinion groups are constructed by users, who are interested in same entities with similar polarities.
Tab. \ref{tab:dataset} sums up basic statistics on the two datasets.
 
\begin{table}[h]
\centering
\vspace{-0.5cm}
\scriptsize
\caption{Statistics of datasets and features we define.}
  \label{tab:dataset}
  \begin{tabular}{lllll}
  \\\hline
  				&&ImagiWeb opinion dataset && RepLab 2013 \\\hline
  source			&&Political opinion tweets 	&& English \& Spanish opinion tweets\\
  annotation size	&&$11527$ tweets ($7283$  unique)	&& $26709$ tweets (all unique)\\
  subsets		&& Entity (politician P, politician Q)		&& Domain (automative, music) \\
  feature space	&& Aspect-polarity pairs				&& Entity-polarity pairs\\
  				&& $9$ aspects, $6$ polarities 		&& $20$ entities per domain, $3$ polarities\\
  \hline\vspace{-0.5cm}
  \end{tabular}
\end{table}

\subsection{Evaluation Measures}
The ground truth is hardly available when evaluating clustering output for evolutionary clustering.
We instead develop the following three quantitative measures with the object of well detecting clustering quality. 

\textit{\textbf{Co-occurrence level.}}
Our main interest lies in detecting compact clusters, which means that the number of observed co-occurring features actually match the estimated distribution.
This measure counts the real number of co-occurring feature couples in each sample among the non-zero features grouped in a cluster.

\textit{\textbf{Unsmoothness.}}
This catches the dissimilarity between corresponding clusters through different time epochs using Kullback-Leibler (KL) divergence.
If a temporal clustering method well detects the evolution of clusters, the cluster signatures having same identity would be similar to each other. Therefore we develop `unsmoothness' to measure how suddenly a cluster changes over time. 

\textit{\textbf{Homogeneity.}}
This measures the degree of unanimity of grouped tweets in a cluster in terms of polarity. Opposite opinions hardly co-occur because an author usually keep his opinion stance in a sufficiently short time. By ignoring the degree of polarity, the homogeneity of a cluster is simply defined as follows\footnote{$\#(\mathrm{polarity})$ is the number of tweets annotated with this polarity.}:\vspace{-0.3cm}
\begin{equation*}
\text{Homogeneity} = {(|\#(\mathrm{positive}) - \#(\mathrm{negative})|)/}{(\#(\mathrm{positive})+\#(\mathrm{negative}))}\vspace{-0.2cm}
\end{equation*}
This is intuitive and easy to be visually represented but is an indirect evaluation.

\subsection{Result}
Clustering is conducted at subset level.
For a given clustering method and subset, experiments are repeated 10 times by changing initialization to get the statistical significance. 
Since MM and PLSA are time-independent, temporal clusters are obtained via two stages: normal clustering per epoch and heuristic matching between clusters of two adjacent epochs judged by their distributional form.

\begin{figure}[h]
 \scriptsize
  \centering
  \vspace{-0.3cm}
  \begin{tabular}{c}
   \includegraphics[width=10cm, height=4.6cm]{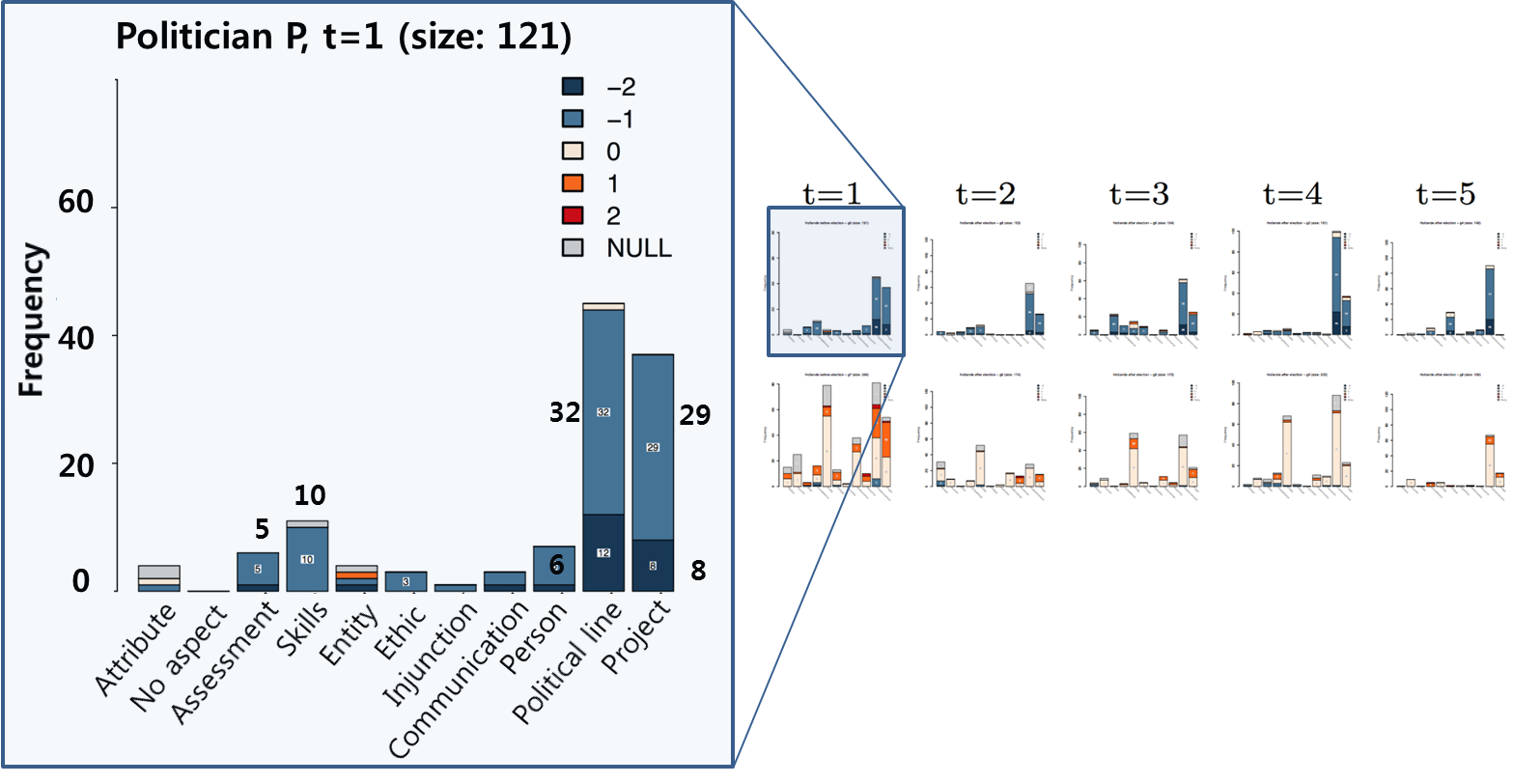}
  \end{tabular}
  \vspace{-0.5cm}
  \caption{Visualization of the evolution of two clusters extracted from a TMM clustering result with five different time epochs on politician $P$ subset.}
  \label{fig:dynamics}
  \vspace{-0.5cm}
\end{figure}

The first sub-table of Table \ref{table:eval} shows the experimental results of four methods on the ImagiWeb dataset. Once clustering is done per subset, we merge the results to analyze together the reputation of two competitors. The number of epochs is fixed at two by splitting data by an actual important political event date. 
Each value is the averaged result of 10 experiments as well as the standard deviation in brackets.
The bold number indicates the best result among four methods and the underlined one is the second best. The gray background of bold number means the result statistically outperforms the second best and the light-gray means it does not outperform the second best, but does the third one. The value of $\alpha$ in TMM has been set to $0.7$ after several pre-experiments judged by visual representation of clusters (as shown in Fig. \ref{fig:dynamics}) as well as balance among cluster sizes. We manually choose the value by varying $\alpha$ from $0.5$ to $1$. Larger value increases distributional similarity whereas decreases separation of opposite opinions. The hyper parameters of DTM have also been set to the best ones after several experiments.

Globally, TMM outperforms the others in terms of two measures except unsmoothness. Then DTM and MM are in the second place. PLSA produces the worst result for all measures. Since homogeneity is a direct basis to evaluate if the tested method well detects the difference between negative and positive opinion groups, it becomes more important when the mix of opposite opinions is a crucial error. Co-occurrence level also directly shows if the captured clusters are really based on the co-occurring features. Given that both measures evaluate cluster quality of a specific time epoch, it is encouraging that TMM provides identical or even slightly better result than MM because TMM can be thought of as a relaxed version of MM in the point of view of data adjustment over time. The result therefore demonstrates that TMM successfully makes use of the generative advantage of MM. For homogeneity, TMM and MM both obtain $0.86$, which perfectly outperform the second best DTM in terms of Mann-Whitney test with the p-value of 0.00001. Meanwhile, for unsmoothness the best one is DTM with a clearly better result, $1.57$ than the others. DTM concentrates on the distribution adjustment over time at the expense of well grouping opinions that is the principal objective in the task. The second best TMM also perfectly outperforms MM with the p-value of 0.0002. It proves the time dependency encoded in TMM successfully enhances MM for capturing cluster evolution.

\begin{table}[t!]
\scriptsize
 \caption{Evaluation of temporal clustering for four methods on ImagiWeb opinion dataset(left) and RepLab 2013 for automative(middle) and music(right).} 
 \label{table:eval}
 \centering
 \begin{tabular}{ccc}
 \begin{tabular}{l||cccc|}
 	\hline
	&\multicolumn{4}{c|}{ImagiWeb opinion dataset}\\
									&\texttt{TMM} 						& \texttt{DTM} 					& \texttt{MM} 					& \texttt{PLSA}\\\hline
	Avg. Homogen.					&\colorbox{gray}{\textbf{0.86}}		&\underline{0.70}				&\colorbox{gray}{\textbf{0.86}}	&0.67\\
	{\tiny (stand. deviation)}			&{\tiny (0.02)}						&{\tiny (0.06)}					&{\tiny (0.02)}					&{\tiny (0.05)}\\
	Co-occurr. level					& \colorbox{light-gray}{\textbf{123}} 	&113							&\underline{122}					&111\\
	{\tiny (stand. deviation)}			&{\tiny (1.98)}						&{\tiny (1.02)}					&{\tiny (0.88)}					&{\tiny (1.48)}\\
	Avg. Unsmooth.					& \underline{2.27}					&\colorbox{gray}{\textbf{1.57}}	& 3.16							&3.61\\
	{\tiny (stand. deviation)}			&{\tiny (0.23)}						&{\tiny (0.10)}					&{\tiny (0.33)}					&{\tiny (0.21)}\\
	\hline
 \end{tabular}
 &
 \begin{tabular}{|cccc|}
 	\hline
	\multicolumn{4}{|c|}{RepLab(Auto)}\\
	\texttt{TMM} 						& \texttt{DTM} 					& \texttt{MM} 				& \texttt{PLSA}\\\hline
	\colorbox{light-gray}{\textbf{0.76}} 	&0.67							&\underline{0.73}			&0.70\\
	{\tiny (0.02)}							&{\tiny (0.05)}					&{\tiny (0.03)}				&{\tiny (0.04)}\\
	\colorbox{gray}{\textbf{ 40 }} 			&\underline{34}					&\colorbox{gray}{\textbf{40}}	&33\\
	{\tiny (1.21)}							&{\tiny (1.18)}					&{\tiny (0.58)}				&{\tiny (1.52)}\\
	\underline{4.30} 						&\colorbox{gray}{\textbf{1.37}}	& 6.35 						&6.91\\
	{\tiny (0.90)}							&{\tiny (0.12)}					&{\tiny (0.82)}				&{\tiny (0.69)}\\
	\hline
 \end{tabular} 
 &
  \begin{tabular}{|cccc}
 	\hline
	\multicolumn{4}{|c}{RepLab(Music)}\\
	\texttt{TMM} 						& \texttt{DTM} 					& \texttt{MM} 				& \texttt{PLSA}\\\hline
	\colorbox{light-gray}{\textbf{0.77}}		&0.75							&0.75						&\underline{0.76}\\
	{\tiny (0.03)}							&{\tiny (0.05)}					&{\tiny (0.02)}				&{\tiny (0.03)}\\	
	\colorbox{light-gray}{\textbf{ 26 }}		&22								&\underline{25}				&22\\
	{\tiny (0.74)}							&{\tiny (0.80)}					&{\tiny (0.40)}				&{\tiny (0.35)}\\	
	\underline{4.5}						&\colorbox{gray}{\textbf{2.54}}	& 6.12						&7.75\\
	{\tiny (0.90)}							&{\tiny (0.51)}					&{\tiny (0.87)}				&{\tiny (1.11)}\\	
	\hline
 \end{tabular}
 \end{tabular}
 \vspace{-0.4cm}
\end{table}

In addition to the quantitative evaluation, we visualize a TMM clustering result in Fig. \ref{fig:dynamics}. It is the evolution of two clusters with five different time epochs on politician $P$ subset. The zoomed figure shows a negative group about $P$ at epoch 1 especially on the aspects ``political line'' and ``project''. TMM captures the dynamics of the cluster over time as shown in the figure. As time goes by, opinions about ``project'' disappear (at t=5) but the other negative opinions about ``ethic'' appear in the cluster. The cluster in the second line groups mainly positive and neutral opinions about various aspects at epoch 1, but some aspects gradually disappear with time. 

The experimental results on RepLab 2013 corpus are given in the middle and right sub-tables in Table \ref{table:eval}. Number of epochs is also fixed at two and the data is split by the median date. 
This corpus is not originally constructed for opinion aggregation, therefore we do not have sufficient feature co-occurrences.
The proportion of instances having at least two components is only $5.2\%$ for automative and $2.9\%$ for music. Despite the handicap, we rather expect that we would emphasize the characteristics of each model via experiments with this restrictive dataset. The $\alpha$ value has been set to $1$ to make maximum use of the effect of previous clusters regarding lack of co-occurrences.

Two outstanding methods are TMM and DTM but there is an obvious difference between their results. TMM gives a better performance in terms of local clustering quality such as homogeneity and co-occurrence whereas DTM outperforms  the others in temporal view. 
Homogeneity does not seem really meaningful here because the opposite opinions about different entities can be naturally mixed in an opinion group. 
However, from the fact that co-occurring features are rarely observed and, moreover, only 10\% of total opinions are negative in the corpus, negative and positive opinions seldom co-occur. 
Therefore, the high homogeneity can be a significant measure here also. 
As in the ImagiWeb dataset, the co-occurrence level of TMM is clearly better than that of DTM. 
On the other hand, even though DTM gives a perfectly better result for unsmoothness, the captured distributions are not really based on the real co-occurrences when we manually verify the result. 
Nevertheless, when the dataset is extremely sparse as in this case, smoothing distribution would anyway provide the opportunity not to ignore rarely co-occurring features. 

\section{Conclusions}
The proposed TMM model succeeds in effectively extending MM, by taking into consideration the temporal factor for clustering. 
Our method captures the dynamics of clusters much better than the heuristic matching of single clustering results using MM or PLSA, without losing clustering quality at local time epoch. 
TMM clearly outperforms DTM in terms of local cluster quality. 
DTM tends to produce well-smoothed distributions over time, but as shown through its low performance with the other measures, high smoothness does not always signify that the cluster evolution is well detected.

An inherent hypothesis in TMM is that clusters evolve progressively over time and it has enabled the modeling of direct dependency between two adjacent epochs. 
However if abrupt changes arrive, the distributions found for each cluster can be incoherent. 
A future developmental direction is taking such changes into account. 
A possible way could be to establish an automatic adjustment of the dependency rate $\alpha$. 
Another interesting direction is to develop means to infer more exactly the conditional probability $p(z^t_{i} | w^t_{im}, z^{t-1}_i)$.

\vspace{-0.4cm}
\subsubsection*{Acknowledgments.} This work was funded by the project ImagiWeb ANR-2012-CORD-002-01.

\vspace{-0.2cm}
\bibliographystyle{splncs}
\bibliography{tmm}   

\end{document}